\documentclass[]{spie}  

\usepackage{amsmath,amsfonts,amssymb}
\usepackage{graphicx}
\usepackage[colorlinks=true, allcolors=blue]{hyperref}

\title{Prototyping the Next-Generation
Spectrograph for Space Exploration: Digital Micromirror Device-based Multi-Object Spectrograph (DMD-MOS) Calibration and Performance Assessment}

\author[a,c]{Jiayi Emma Xu}
\author[a,b]{Sarik Jeram}
\author[b]{Shaojie Chen}
\author[a,b]{Ting S. Li}
\author[a]{Ziang Chen}
\author[d]{Zain Azam}
\author[e]{Adi Khandelwal}
\author[b]{Mark Barnet}
\author[a]{Yixuan Cheng}
\author[a,b]{Suresh Sivanandam}

\affil[a]{David A. Dunlap Department of Astronomy and Astrophysics, University of Toronto, 50 St.George, Toronto, Canada}
\affil[b]{Dunlap Institute for Astronomy and Astrophysics, University of Toronto, 50 St.George, Toronto, Canada}
\affil[c]{Department of Physics, University of Toronto, 65 St.George, Toronto, Canada}
\affil[d]{Department of Electrical and Computer Engineering, University of Toronto, 50 St.George, Toronto, Canada}
\affil[e]{Steward Observatory \& Department of Astronomy, University of Arizona, US}

\authorinfo{Jiayi Emma Xu: \\E-mail: emmajiayi.xu@mail.utoronto.ca, \\Personal website: \url{https://acedemic-website.vercel.app/}}

\begin{document} 
\maketitle

\begin{abstract}
    Abandoning the traditional approaches to multi-object spectrographs, the Digital Micro-mirror Device-based Multi-Object Spectrograph (DMD-MOS) leverages a robust, elegant method for creating multi-object spectroscopy slit masks through use of a DMD, which is an array of microscopic flat mirrors where each micro-mirror can be addressed to physically flip between two discrete angles. We have designed two optical channels for DMD-MOS - one for each of these two positions: 1) an imaging channel that can be used for multi-object target selection and 2) a spectrograph channel. Using this two-channel approach, DMD-MOS can simultaneously image the telescope field of view (FOV) with the imaging channel and actuate the specific DMD micro-mirrors to send light from chosen targets into the spectrograph channel. In this work, we present the results of our laboratory testing of DMD-MOS including spectral resolution, line spread function, wavelength coverage, and the mapping between the DMD and imaging channel planes, the latter of which enables DMD-MOS and our control software to instantaneously configure the DMD to observe chosen target(s) in the acquisition field.
\end{abstract}


\section{Introduction}
\label{sec:intro}  


In the past few decades, there have been many developments in astronomical spectrographs to meet the needs of modern astronomical research. To better use observation time and study interacting stars or stars in a cluster, multi-object spectrographs (MOS) were developed. MOS breaks the limitations of older generation spectrographs by allowing multiple spectra to be obtained in a single pointing, thus significantly increasing the observation efficiency, enabling science cases like stellar surveys. Current popular multi-object spectrographs are fiber-based or use fabricated slit masks. Fiber based MOS such as Gemini (GMOS) [\citenum{allingtonsmith2002gmosintegralfieldunit}] applied movable fibers at the focal plane for each target's location, with the light from each target independently propagating to the detector. This requires a large image panel that allows the fibers to fit [\citenum{10.1117/12.3019896}]. Fabricated slit mask-based MOS, on the other hand, involves cutting a slit mask out of metal at the targeting location using high precision laser cutting methods. This process involves target selection, making extremely precise measurements and predictions, mask manufacturing, installation, and then finally observing, taking as long as 3-4 weeks from start to finish. Motivated by the need for small and simple MOS spectrographs, DMD-MOS was born.

Digital Micro-mirror Devices (DMDs) are high precision opto-mechanical instruments with an array of micro-mirrors. Each micro-mirror can actuate between two angles at $\pm12$ degrees and can be individually controlled by computer. Furthermore, the response time of each mirror is within 50 $\mu s$ [\citenum{10.1117/1.JMM.13.1.011104}]. Given these properties, a DMD can be used as an effective, quickly programmable MOS slit mask - a concept explored by instruments like RITMOS [\citenum{10.1117/12.549897}] on the C.E.K. Mees telescope, SAMOS on the SOAR telescope [\citenum{10.1117/12.2311876}, \citenum{10.1117/12.3020796}] and D-MOS for INdian Spectroscopic and Imaging Space Telescope (INSIST) [\citenum{2025ExA....60...16S}]. The above ground-based spectrographs were the pioneers of DMD-based visible and NIR MOS that demonstrated the feasibility of the concept. Beyond these, a few space-based missions such as SUMO [\citenum{10.1117/12.2594651}], EUCLID [\citenum{10.1117/12.825641}], and SIRMOS [\citenum{10.1117/12.3017865}] have broadened the possibility of such designs for spaced-based UV science. Travinsky's et al. [\citenum{10.1117/1.JATIS.3.3.035003}] characterized the robustness of DMDs and report 0 stuck micro-mirrors after 200,000 flips and $99.62\%$ mirror flipping accuracy after radiation, vibration, shock, and low temperature tests, concluding DMDs to be highly reliable devices that meet the standards for space missions.

Similarly to some of the instruments mentioned above, our design features two optical channels: 1) the imaging channel for field of view observation and target selection and 2) the spectral channel. Unlike those projects, DMD-MOS occupies a small footprint of approximately $20\times30\times40$cm on a trapezoidal optical bench along with a handle at the top of the enclosure, facilitating transportation for use at multiple observatories. The spectral channel is a volume phase holographic (VPH) grism-based spectrograph with a spectral resolution of 1000 at 550 nm when using a 2$\times$2 DMD micro-mirror slit. As mentioned previously, DMD-MOS is able to use both of these channels for science observations. This process starts by astrometrically solving the imaging channel field after which the imaging-channel-to-DMD mapping is used to flip the mirrors corresponding to desired targets toward the spectral channel. Science targets are specified with a user-provided list of sky coordinates but manual target selection is also possible. The raw data products generated by DMD-MOS can then be reduced through our data reduction pipeline.


This work aims to quantify the performance of both channels. In Section \ref{sec:design}, we provide a brief introduction of the system design as well as the test bench. In Sections \ref{sec: imaging channel} and \ref{sec: performance}, we discuss the performance of the imaging and spectral channels, respectively. Section \ref{sec:mapping} discusses the mapping relationship between the DMD and imaging channel detector as well as the mapping accuracy. After concluding lab testing, we performed on-sky tests, the results of which we discuss in Section \ref{sec: sky test}. Lastly, we discuss future DMD-MOS work in Section \ref{sec: future work} and conclude in Section \ref{sec:conclusion}.

\section{System Design and Testing Method}\label{sec:design}

\begin{figure}[htb!]
    \centering
    \includegraphics[width=0.7\linewidth]{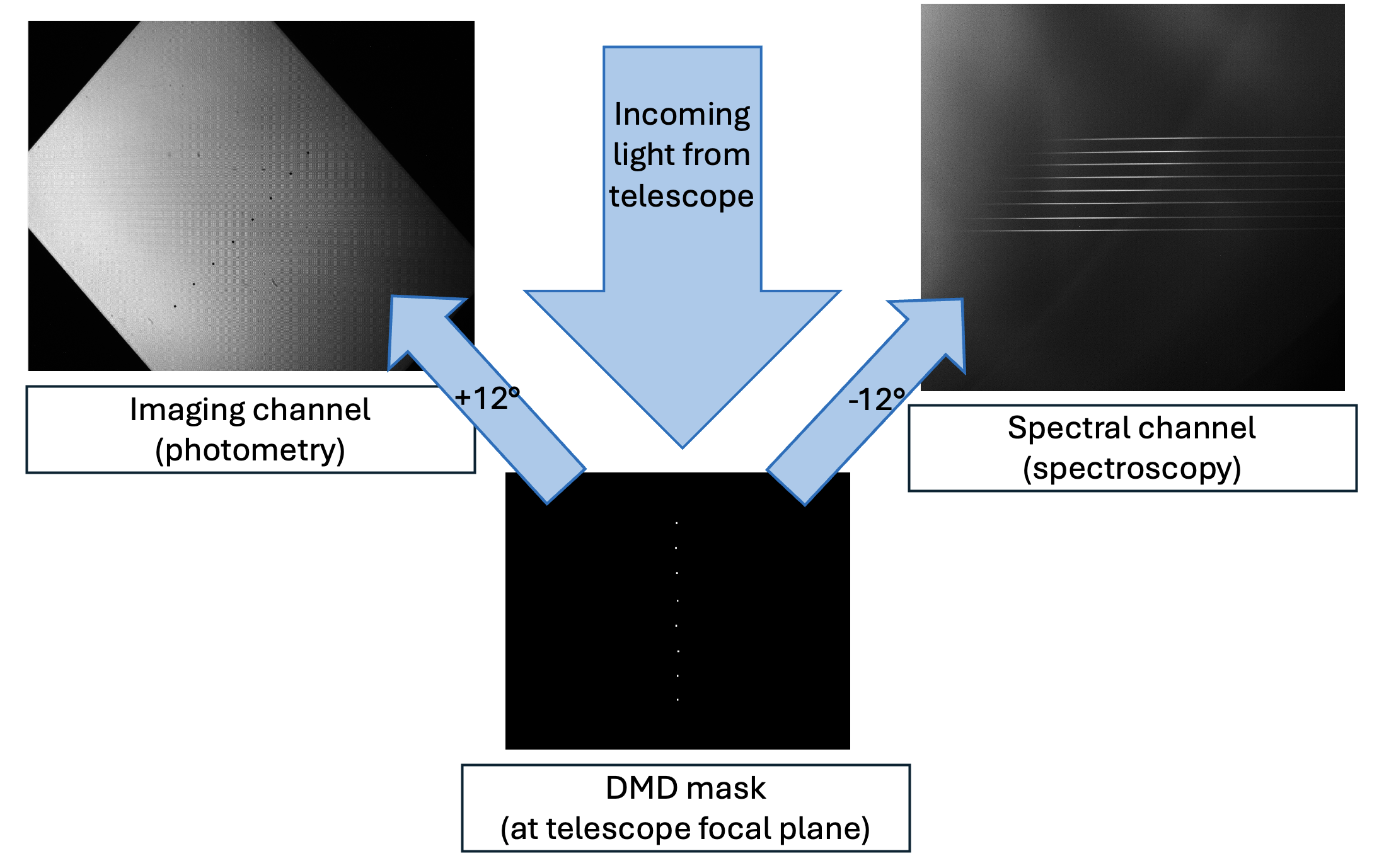}
    \caption{Idea of DMD-MOS design with real image captured from the Imaging channel, Spectral channel using the mask showed.}
    \label{fig:idea}
\end{figure}

DMD-MOS contains 3 main components: the Digital Micro-mirror Device (DMD), the imaging channel, and the spectral channel. We use the DMD micro-mirrors' $\pm12$ degree actuation to separate incoming light into these latter two optical channels. This is depicted in Figure \ref{fig:idea}, where we use a mask with 8 DMD `pinhole' slits.
Note that the DMD mask uploaded to the DMD is a large binary array - a value of 0 flips the corresponding micro-mirrors towards the imaging channel (+12 degrees), while a value of 1 flips them towards the spectral channel (-12 degrees).
After the input light is sent towards the spectral channel, we observe the expected spectra corresponding to each of the 8 slits while also rejecting extraneous light from background sources.

\begin{figure}[htb!]
    \centering
    \includegraphics[width=0.65\linewidth]{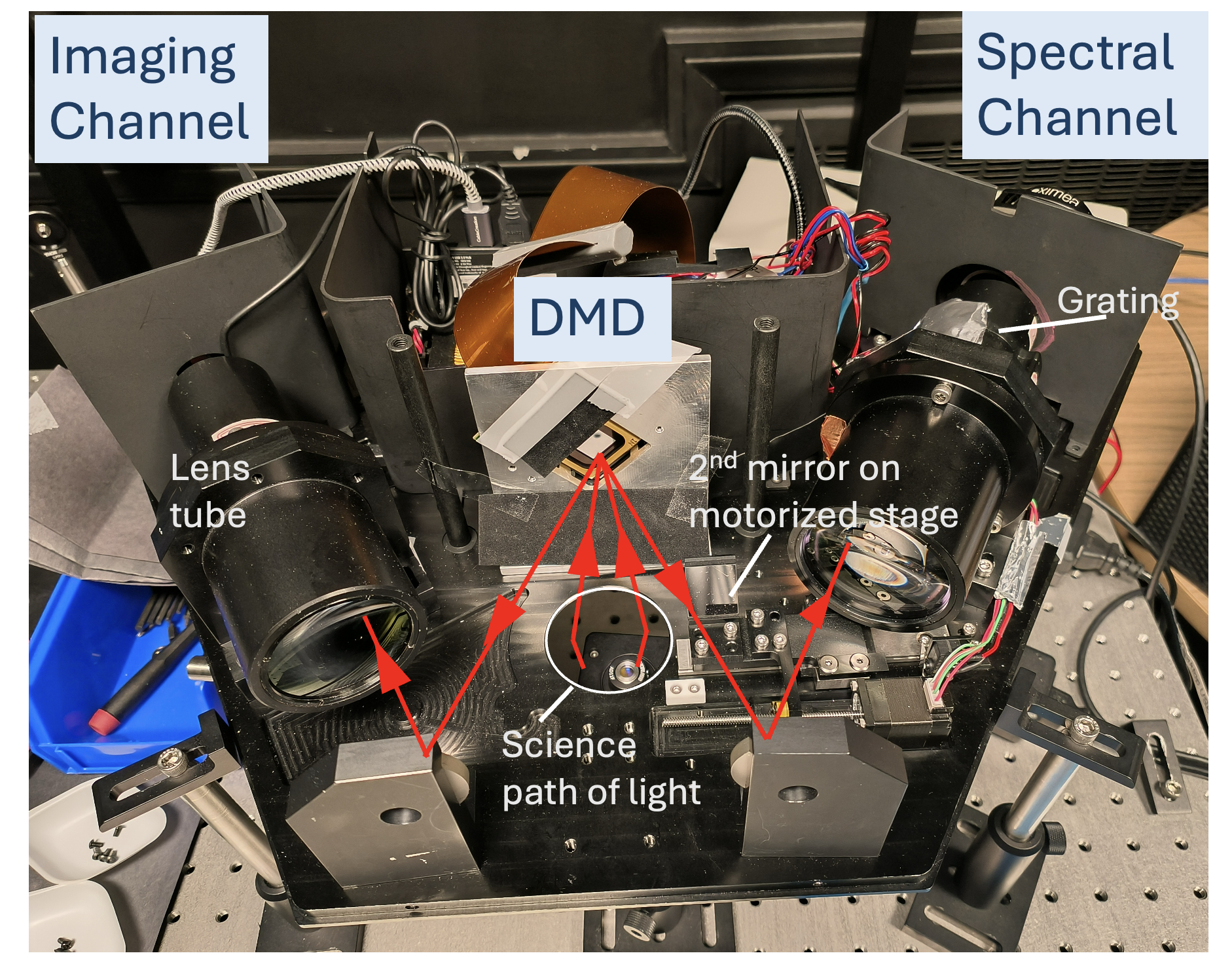}
    \caption{Inner view of DMD-MOS. DMD, Imaging channel and Spectral channel labeled in blue, optical component labeled in white, light path labeled by red line with arrow indicating direction. The applied DMD with corresponding Image and Spectrum were also showed in the figure.}
    \label{fig:DMD_design}
\end{figure}

In Figure~\ref{fig:DMD_design}, we show DMD-MOS with the enclosure removed along with the optical path of the input from the telescope. For our DMD, we use a Texas Instruments DLP7000. For our detectors, we use Ximea XiC and XiJ series detectors for the imaging and spectral channels, the latter of which uses a thermoelectrically cooled (TEC) scientific CMOS (sCMOS) image sensor. 
Focused light from the telescope enters through an opening at the bottom of the optical bench (we will refer to this as the ``science path") that is then incident on a fold mirror (``science fold") that directs the light towards the DMD. The DMD will then selectively send light into the imaging and spectral channels by use of fold mirrors in each channel. The imaging channel features a collimator and camera to image the DMD plane. The spectra channel features a collimator, VPH grism, and camera; by using a grism, we are able to output the $m=1$ spectrum order at 0 degrees, keeping the overall footprint of DMD-MOS compact.


Beyond the DMD and two optical channels, DMD-MOS features a calibration sub-system consisting of a secondary mirror (``calibration fold") on a motorized stage and a fiber collimator. A fiber collimator sits at the base of the science fold; by moving the calibration fold into position, the light from the fiber collimator is then able to be directed on to the DMD, allowing for fiber-fed arc lamp and continuum sources to illuminate the slits that are used for obtaining science data. Note that due to the actuation axis of the DMD micro-mirrors, the DMD must be placed at a 45 degree angle as seen through the imaging channel image in Figure ~\ref{fig:idea}. 
This been said, each micro-mirror is physically orientated as a diamond, which may cause uneven amount of light entering the spectral channel. Thus, it is worthy to evaluate different slit masks and examine the point spread function from those slits.
Looking at the entire field of the Imaging channel, as a consequence of DMD orientation, the corners of the DMD are cut off, ultimately reducing the total usable area of the DMD. For more details on the design, we refer the reader to Shaojie's paper [\citenum{10.1117/12.2630372}].

\begin{table}[htb!]
    \centering
    \begin{tabular}{|l|l|l|}
    \hline
        \textbf{Components}& \textbf{Parameters} & \textbf{Values}\\\hline
        DMD & Resolution & $1024\times768$ \\
        &Micro-mirror size & 13.7$\mu$m \\\hline
         Imaging channel& Detector resolution & $2064\times1544$\\
         {}& Pixel size & 3.45$\mu$m \\
         {} & Magnification & 0.5 \\
         {} & Image quality & EE50@ $4\times4$ px \\ \hline
         Spectral channel &Detector resolution & $2048\times2048$\\ 
         {}& Pixel size &6.5$\mu$m \\
         {}& Wavelength range & 0.4-0.7$\mu$m \\
         {}&Spectral resolving power  & \~1000  \\ 
         {} & Magnification & 0.47\\
         {} & Image quality & EE50@ $2\times2$ px\\\hline
         
    \end{tabular}
    \caption{Basic parameters for DMD-MOS. Full table see Table 1 in [\citenum{10.1117/12.2630372}].}
    \label{tab:system parameters}
\end{table}
Table~\ref{tab:system parameters} shows the main parameters of DMD-MOS. The imaging channel and spectral channel are resolvable as they are about 2 times bigger than the DMD arrays, so an approximate correspondence is that 1 micro-mirror corresponds to $2\times2$ imaging channel detector pixels.

For optical alignment, calibration, and preliminary testing, we used an autocollimating alignment telescope (AT) to project collimated light directly onto the DMD. Though this does not follow the same optical path as the science path, which accepts a converging telescope field, this approach allows for obtaining optimal image quality. Ultimately, both collimated and focused beams were used for testing the system and we found no noticeable difference in the output. One of the limitations of using the AT is the beam size - the beam overfills the DMD array, making it difficult to test point source targets (e.g. simulated stars). In addition to the AT, we used several different arc lamps for both alignment and wavelength calibration, the results of which we show in Section~\ref{sec: performance}. 

\section{Imaging channel testing}\label{sec: imaging channel}

During normal use, we use the imaging channel to observe and monitor the telescope FOV. Though it is not currently used for any scientific observations (i.e. photometry), the imaging channel must be able to image the telescope field with sufficient image quality.
In this this Section, we analyze the imaging channel's image quality by fitting a top-hat function to the slit profile, calculating the full width at half maximum (FWHM) using a ``pinhole" mask, and quantifying the resolution.

\begin{figure}[htb!]
    \centering
    \includegraphics[width=0.5\linewidth]{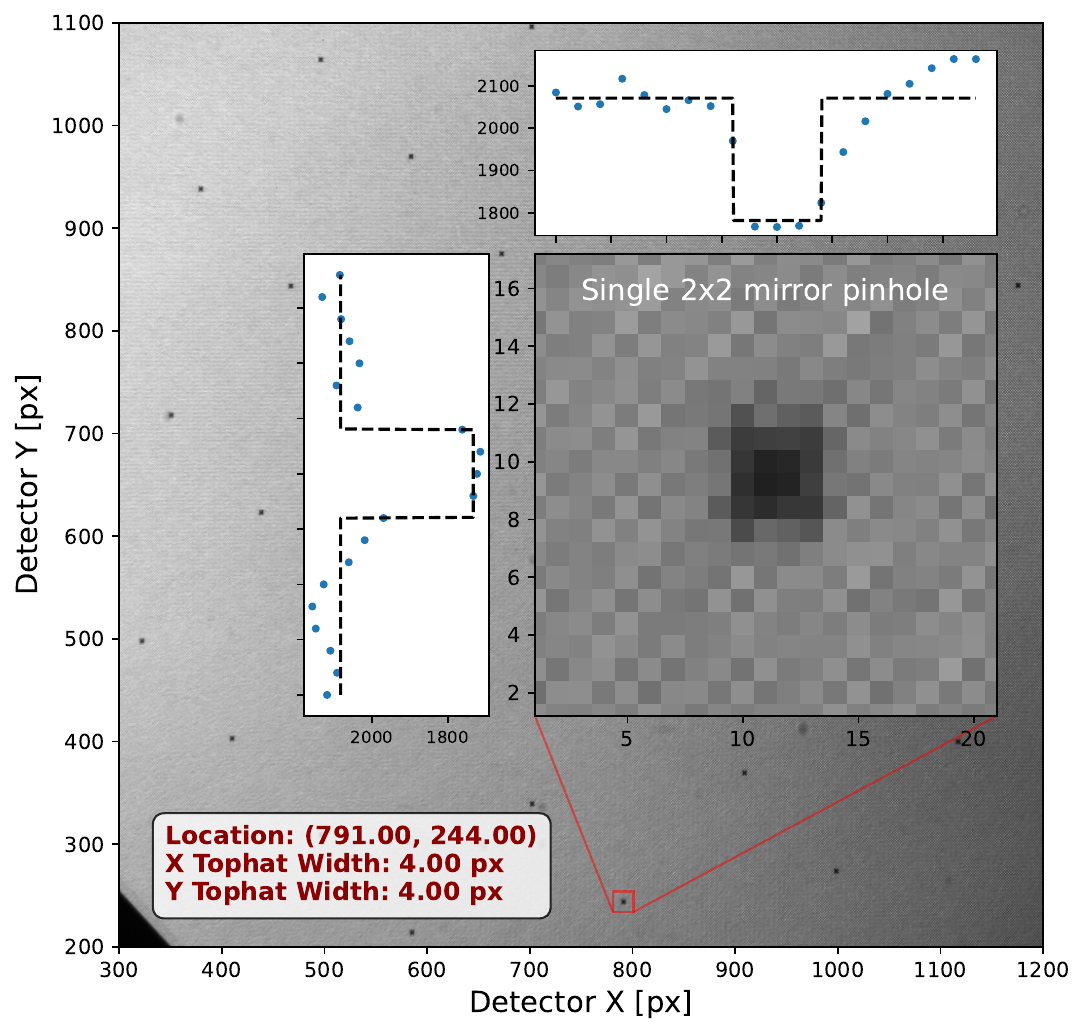}
    \caption{Imaging channel with pinhole mask. Each single pinhole consists $2\times2$ micro-mirrors. Zoomed in pinhole mask (with point-like light source enter spectral channel) with top-hat fit was plot on top of the full mask. }
    \label{fig:pinhole imaging}
\end{figure}
Figure~\ref{fig:pinhole imaging} shows a portion of the imaging channel with an illuminated $10
\times10$ pinhole grid, with each pinhole a $2\times2$ micro-mirror grid directed towards the spectral channel. The inset figure in Figure~\ref{fig:pinhole imaging} shows a close-up of one pinhole; here, the darkest pixels correspond to micro-mirrors directed to the spectral channel. We observe that the edges of these pinholes are relatively sharp, as expected. We fit a top-hat function along both X and Y axes and obtained a top-hat width of approximately 4 pixels in both directions, matching the expected result as two imaging channel pixels correspond to a single DMD in width.

To further test the image quality, we use an inverted version of the pinhole mask, where most of the DMDs are directed towards the spectral channel, resulting in the pinhole spots being directed into the imaging channel. As we no longer expect a top-hat function for the intensity distribution, we instead fit a Gaussian distribution to determine the FWHM, which allows for a direct comparison with the system's Zemax optical model parameters. 

\begin{figure}[htb!]
    \centering
    \includegraphics[width=0.98\linewidth]{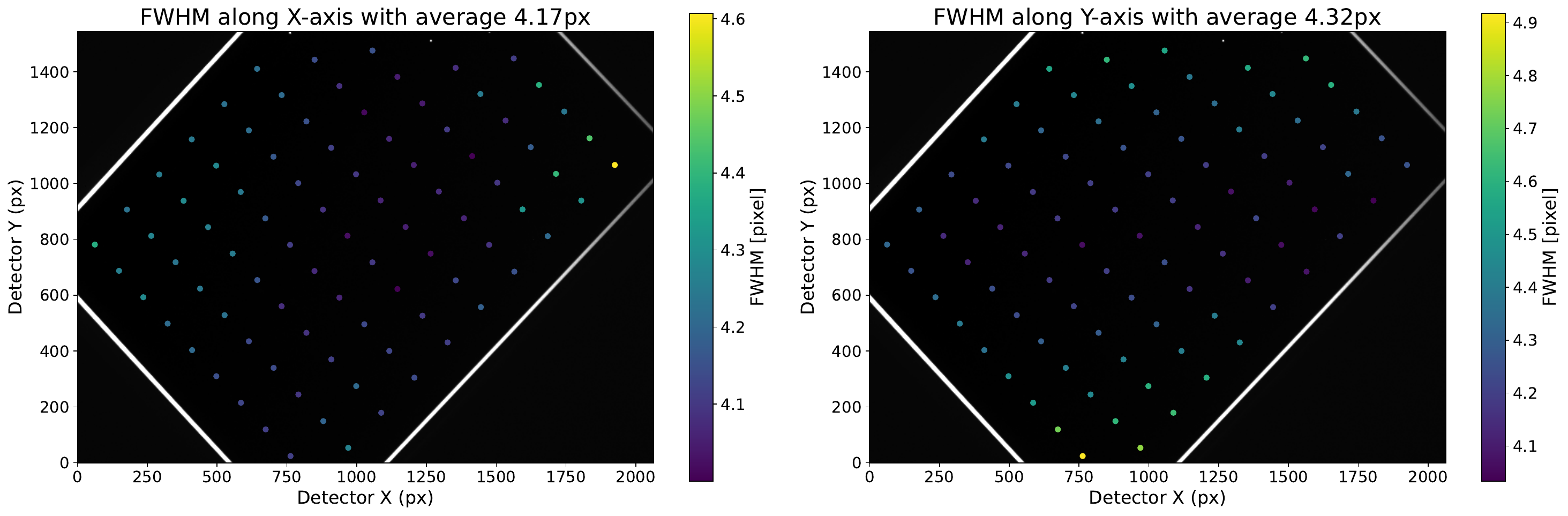}
    \caption{FWHM analysis for Imaging channel with mean value along X-axis (left) and mean value along Y-axis (right) using inverted pinhole mask with light hitting the image channel at the pinholes. Each pinhole is $2\times2$ DMD micro-mirros. Colorbar indicate FWHM in the unit of pixels. }
    \label{fig:FWHM imaging}
\end{figure}
Figure~\ref{fig:FWHM imaging} shows the FWHM analysis for the imaging channel with a $10\times10$ pinhole mask. Each pinhole consists of $2\times2$ micro-mirrors on the DMD. The FWHMs of the pinholes are coded by color for the X-axis (left) and Y-axis (right). We observe smaller FWHMs in the center and larger FWHMs near the edges in both of these plots. Additionally, all FWHMs were bound between 4 and 5 pixels, so the deviations are at the sub-pixel level. The average values of 4.17 pixels along X and 4.32 pixels along Y confirm agreement with our expectation of 1-to-2 micro-mirror-to-imaging-channel-pixels correspondence.

\section{Spectral channel testing}\label{sec: performance}

In normal use, after selecting targets using the imaging channel and directing the corresponding DMDs toward the spectral channel, we obtain the spectrum of the selected targets. To analyze the image quality of the spectral channel and evaluate the wavelength dispersion and spectral resolution, we use pinhole and ``staircase" masks for testing and report our results later in this Section.

\subsection{FWHM Tests with Different Masks}\label{sec:FWHM}

Similar to the imaging channel analysis, we use FWHM analysis on the spectral channel data as it allows for quantifying the intensity distribution and image sharpness, which allows us to check for defocus and astigmatism. We obtain spectra using 1) a pinhole mask, where the emission lines were point-like after dispersion and 2) a segmented long-slit mask, similar to long slits used in spectroscopy.

\subsubsection{Pinhole masks}

We first use a single micro-mirror as the slit with a 60 second exposure time for a Hg light source. Because Mercury has three distinctive emission lines at 546 nm, 577 nm and 579 nm, the former of which is close to our design central wavelength (550nm), we use Hg arc lines for alignment checks as these three emission lines are readily identified.

\begin{figure}[htb!]
    \centering
    \includegraphics[width=0.6\linewidth]{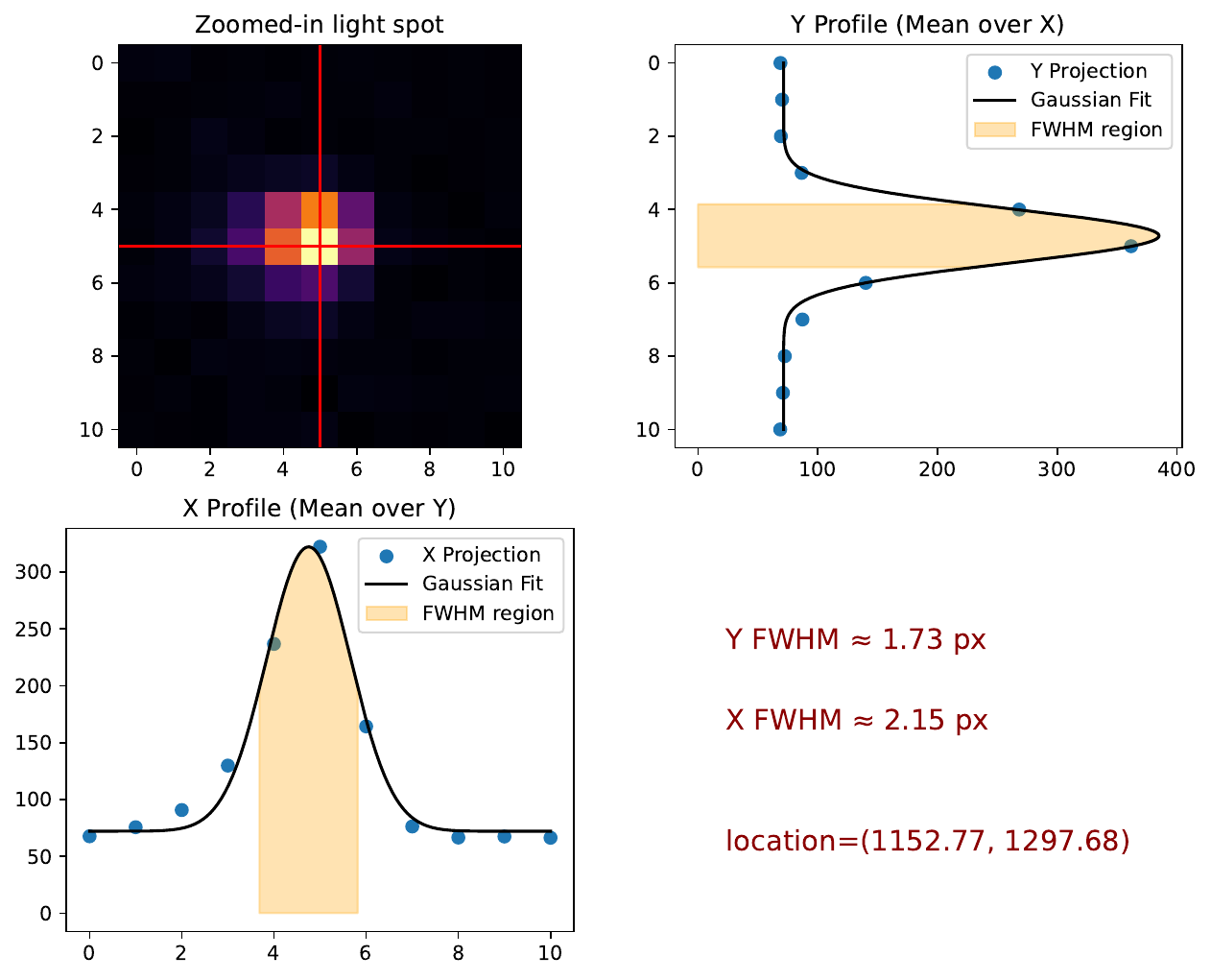}
    \caption{Zoom in of the 546nm Hg emission line with a one micro-mirror slit (top-left). PSF analysis of the light spot along Y-direction with average intensity (blue dots), Gaussian fit (solid black line), and FWHM region (light orange) (top-right). PSF analysis along X-direction (bottom left). Calculated FWHM and the location of the spectral line on the spectral channel detector (bottom-right).}
    \label{fig:single pinhole}
\end{figure}

Figure~\ref{fig:single pinhole} shows the resolution of the system where one DMD mirror corresponds to, on average, $2.15 \times 1.73$ pixels in the spectral channel. The blue dots represent the average intensity along the chosen direction, with X corresponding to the spectral direction and Y corresponding to the spatial direction. The PSF in both directions is approximately Gaussian as the black line shows. The region corresponding to the FWHM is labeled in orange and the value is shown on the bottom right panel together with the location of the arc line spot on the spectral channel detector.

To gauge the performance across the entire spectral channel, we generate a new pinhole mask consisting of three vertical dotted line slits, with each dot corresponding to $3\times3$ micro-mirrors on the DMD. Through these three dotted line slits, we observe the previously mentioned Hg emission lines, along with an additional 435 nm line for the middle dotted line slit, resulting in 10 sets of emission lines in total.


\begin{figure}[htb!]
    \centering
    \includegraphics[width=0.98\linewidth]{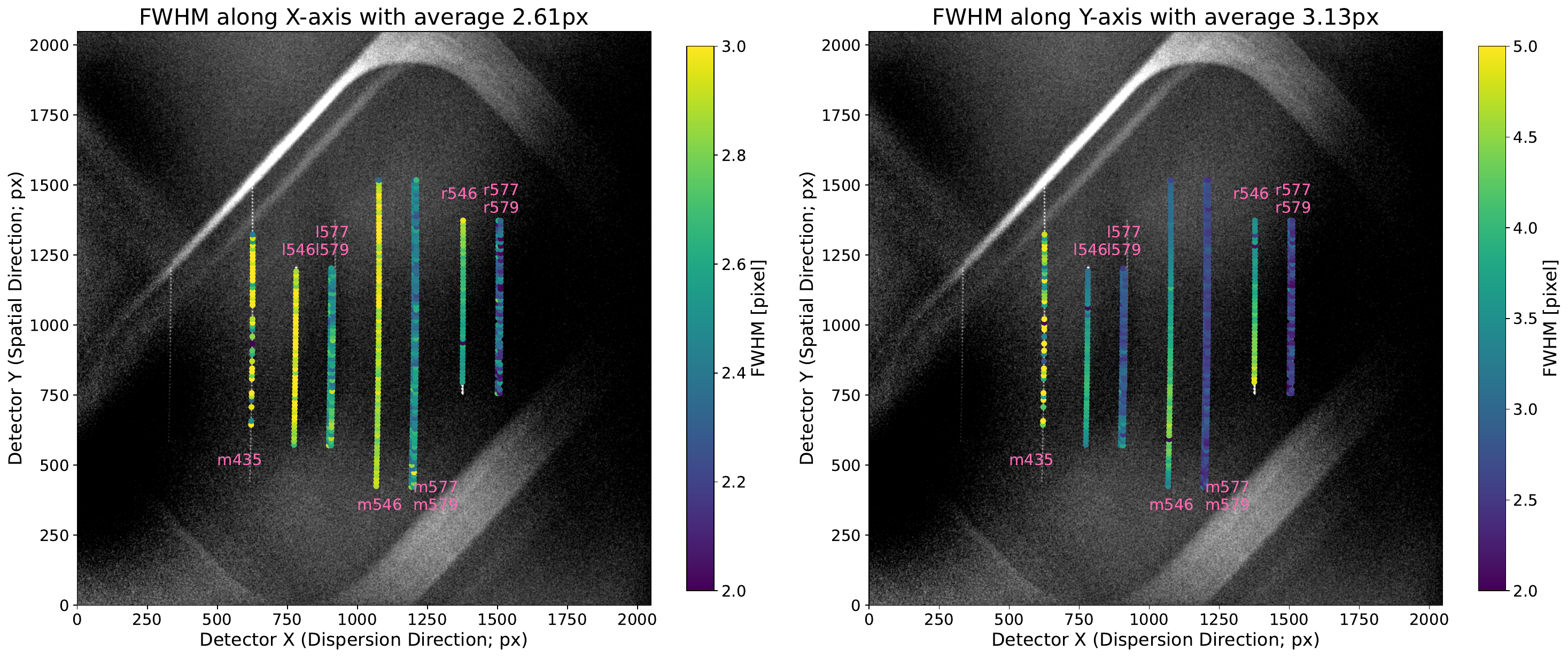}
    \caption{Color coded FWHM calculation along X (left) and Y (right) axis, from Hg emission lines through three vertical dotted line slits. Each dot corresponds to $3\times3$ micro-mirrors flipped to the spectral channel. In the line labels, ``l,m,r" represent left, middle, and right slit, respectively, followed by the wavelength of the line (e.g.``m546" corresponds to the 546 nm Hg line of the middle slit).}
    \label{fig:FWHM of Hg spectrum}
\end{figure}
Figure~\ref{fig:FWHM of Hg spectrum} shows the FWHM of the individual dots of the dotted line slit mask as measured on the spectral channel. 
For a $3\times3$ DMD pinhole, we see that the spectral direction FWHM ranges from 2 to 3 pixels with an average of 2.61 pixels while the spatial direction FWHMs ranges from 2 to 5 pixels with an average of 3.13 pixels. 
For all three slits, the 546 nm line has slightly larger FWHM values compared to the 577 nm and 579 nm lines. The SNR for 546 nm lines are above 50, but the other lines only have SNR less than 5, where the strong background stray light causing weaker identification of the light spots size. Thus, for a more consistent calculation and a more direct comparison with the result from Figure~\ref{fig:single pinhole}, we excluded the low SNR lines, and only kept the 546 nm lines from all three slits for further analysis. The FWHMs we measured for this specific line across the spectral detector is 3.74 pixels in the spectral direction and 2.81 pixels along the spatial direction. This agrees with the expected 1:1 ratio for DMD mirror and spectrum spot size from optical design. As the FWHM is wavelength dependent, the value along spectral direction is slightly bigger, but still within sub-pixel difference range. 

The image quality of the spectral channel depends on both field position and wavelength. Since a given detector location may correspond to different combinations of field points and wavelengths, maintaining consistent image quality across the full FOV and spectral range is a significant design challenge. The optical design was optimized to achieve uniform performance over the entire FOV. For spot diagrams at both central and edge wavelengths across the FOV, we direct the reader to Section 3.3 of the companion paper ``Optical Design and Wavelength Calibration of a DMD-based Multi-Object Spectrograph"[\citenum{10.1117/12.2630372}]. Our laboratory measurements are in agreement with the optical model, confirming that the as-built system achieves the predicted imaging performance.

\subsubsection{Long-slit Masks (Staircase Slits)}
We additionally test the image quality and wavelength calibration when using long-slit masks similar to those that will be used during on-sky observations.

\begin{figure}[htb!]
    \centering
    \includegraphics[width=0.95\linewidth]{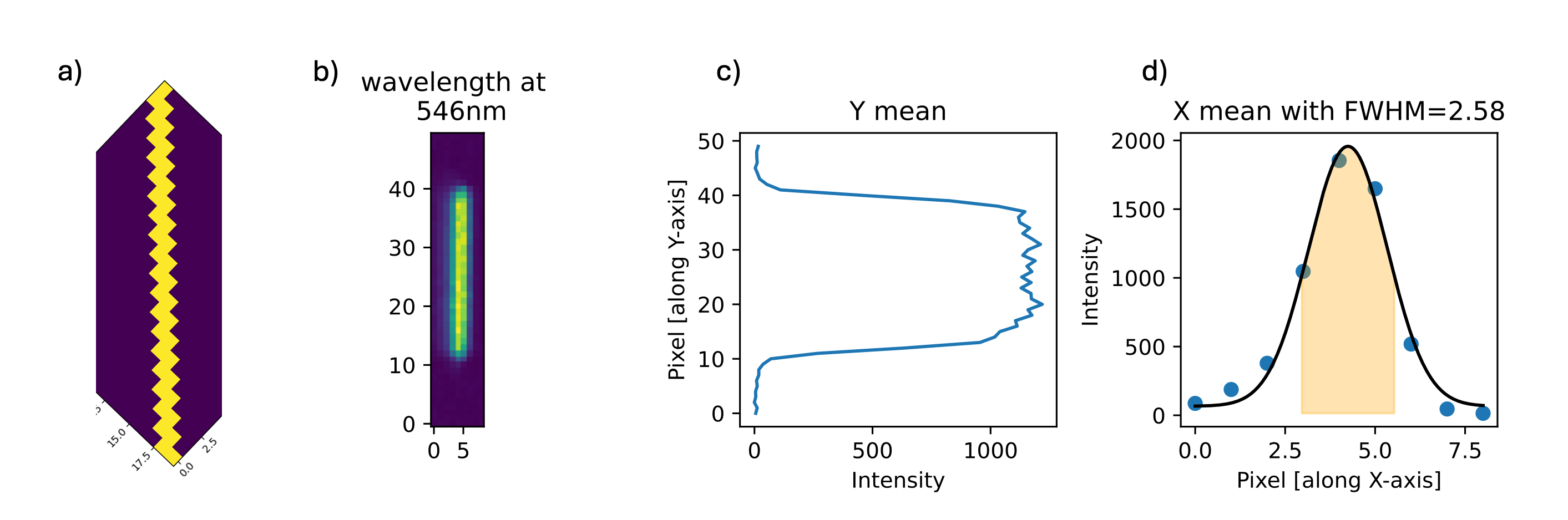}
    \caption{a) Staircase slit with uniform slit width of $\sqrt{2}$ micro-mirrors across a total length of $20\sqrt{2}$ mirrors, with purple and yellow pixels directing DMDs towards the imaging and spectral channels, respectively. This is the minimum width unit for uniform slits, and both length and width are adjustable. b) Emission line at 546 nm from Hg arc lamp through the staircase slit mask. c) Mean intensity along Y-direction of this emission line. d) Mean intensity along X-direction with Gaussian fit (black line) and FWHM range (light orange).}
    \label{fig:stair case slit}
\end{figure}

Due to the geometry of the DMD and the individual micro-mirrors, which are oriented as diamonds rather than squares, more of the light from targets is directed into the spectral channel at the center of a slit compared to the top and bottom of the same slit. To address this problem, we use a ``staircase" slit, as shown in panel a) of Figure~\ref{fig:stair case slit}. We see that this slit has a uniform width along the vertical direction where the horizontal distance of the yellow slit is always $\sqrt{2}$ micro-mirrors wide, corresponding to 1.15 arcseconds on-sky (with a 0.5 m telescope). One concern with using this slit geometry is whether the spectrum from this slit manifests in a staircase-like intensity distribution in the spectrum. In Figure~\ref{fig:stair case slit}, we evaluate one of the emission lines and test if the intensity distribution along the spectral direction is Gaussian using the 546nm line from our Hg arc lamp. 
 Panel c) shows the mean intensity value of the emission line along the spatial direction, where we could clearly identify the signal and the background. The variation of the intensity between Y-axis pixels 10-40 is $\sim$4\% of the total intensity. Panel d) shows the FWHM of the emission line is 2.58 which was labeled in light orange. From the spatial and spectral direction intensity distributions, we see that the PSF of the emission line from the staircase slit follows a Gaussian function and the FWHM value is consistent with the result from the pinhole mask, indicating that the width is indeed generally uniform.

\subsection{Wavelength calibration}\label{sec: wavelength calib}

\begin{figure}[htb!]
    \centering
    \includegraphics[width=0.9\linewidth]{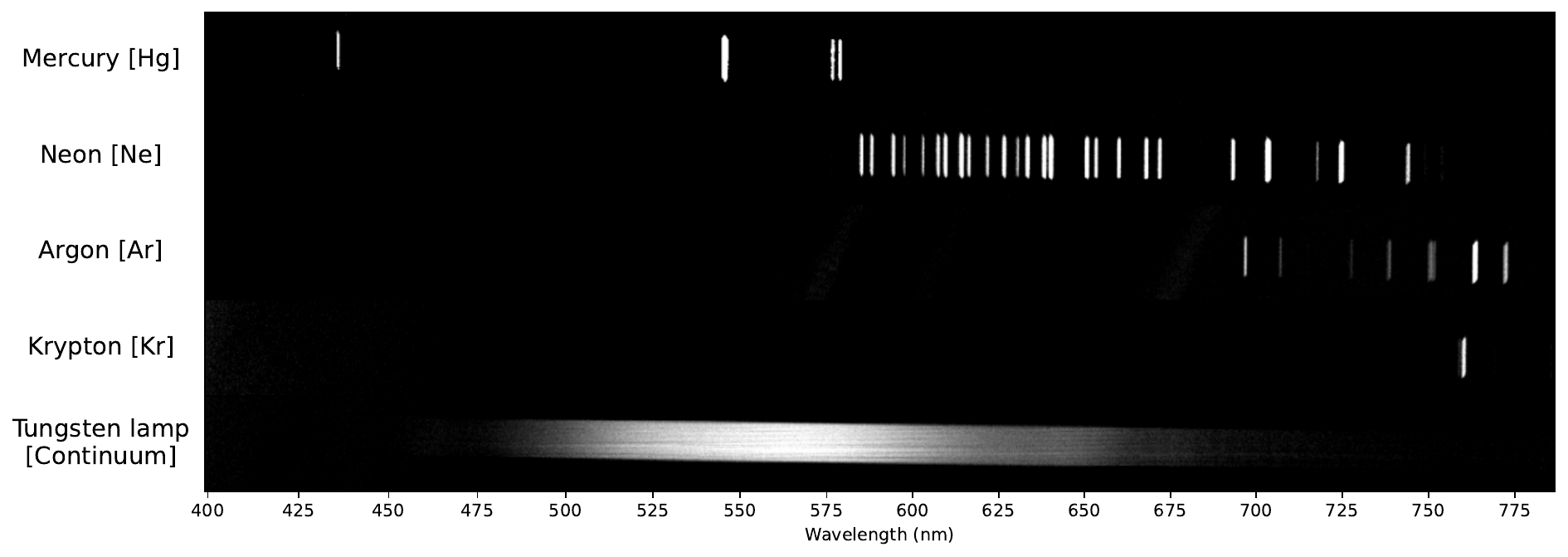}
    \caption{Wavelength-calibrated spectra of different calibration light sources captured using a staircase slit at the center of the DMD. Among these light sources, Hg, Ne, Ar, Kr were used for manual wavelength calibration.}
    \label{fig:Arclamps}
\end{figure}

We use a variety of different light sources for wavelength calibration tests. In Figure~\ref{fig:Arclamps}, we show the spectra of these different sources using the staircase slit from Figure~\ref{fig:stair case slit} placed at the center of the DMD. Four different arc lamps are used: Mercury (Hg), Neon (Ne), Argon (Ar), Krypton (Kr); we additionally use a tungsten lamp as a continuum source. 
By manually matching the known wavelengths of the emission lines to their pixel locations, we are able to generate a wavelength solution. For these lamps, we note that we have more emission lines at the redder end of the spectrum, but few emission lines bluer than 435 nm, which hinders our calibration of the blue end of the spectrum. Using a 3rd-order polynomial fit, we determine that one pixel covers $\Delta\lambda$ of 0.25 nm near the center wavelength (550 nm) at the center of the detector. More than 95\% of the residuals are within 1 pixel with a maximum residual from a single line of $\sim$0.75 nm, corresponding to 3 detector pixels. We note that for science data processing, our data reduction pipeline uses a more robust algorithm for generating the wavelength solution and in laboratory test data, the residuals are typically less than one pixel.

\section{DMD-Imaging mapping relation}\label{sec:mapping}

One of the highlights of DMD-MOS is that an observer can manually select any target in the FOV and the system will generate a slit at the desired location. To achieve this, we must have a mapping between the DMD and the imaging channel detector. We use point-correspondences to map DMD micro-mirrors to their corresponding imaging channel detector pixels, and the correspondences are then input into a solver that outputs the transformation matrix (the homography) that allows the conversion from imaging channel pixels to DMD micro-mirrors.

\subsection{Mapping Method}\label{sec:mapping method}
The homography is a $3\times3$ transformation matrix that map the points of one planar surface onto another surface that may have undergone changes in scale, rotation, shear and flip. Due to the orientation of the DMD, the imaging channel generally only captures the central portion of the DMD (the corners are inaccessible).
\begin{figure}[htb!]
    \centering
    \includegraphics[width=0.9\linewidth]{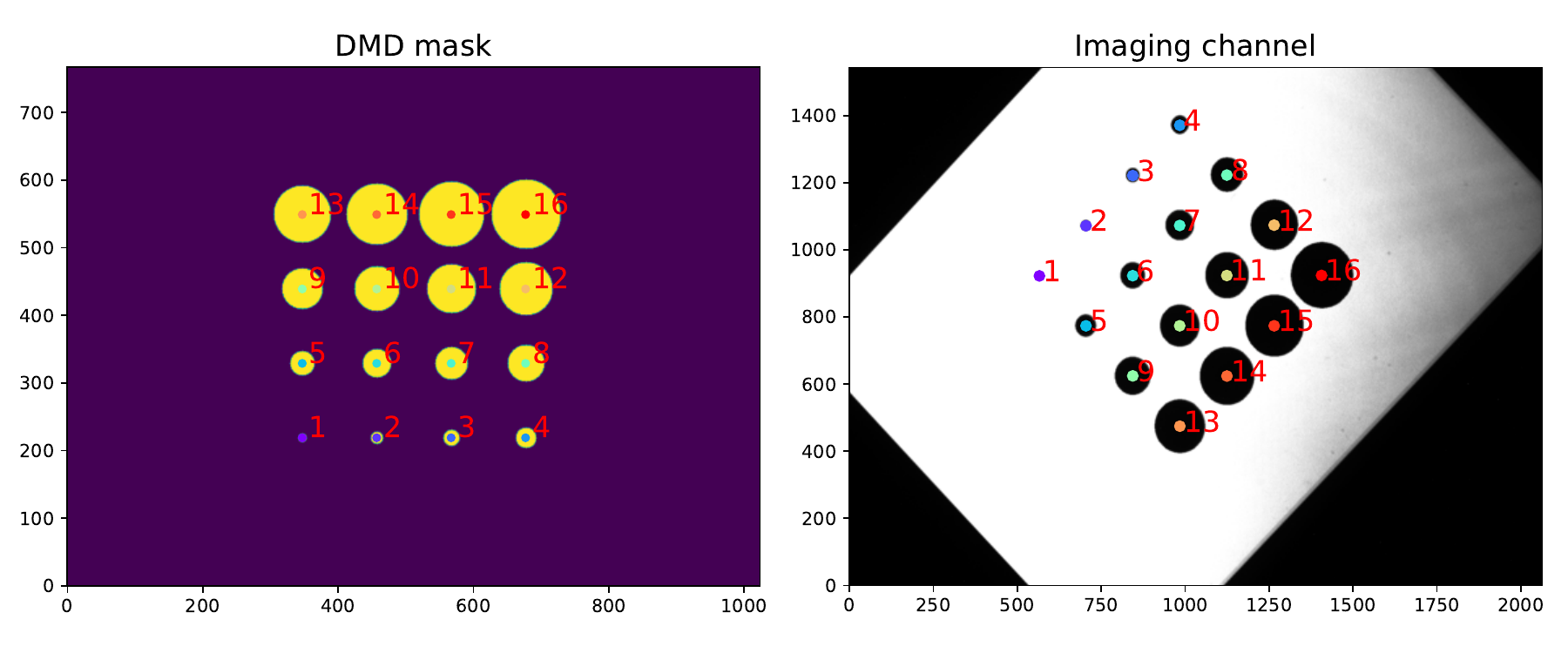}
    \caption{DMD mask with 16 different size circles labeled by the order of increasing area (left). Corresponding imaging channel with the same mask with labeled order (right). The centroids of the circles are labeled by matching color on both sides by matching the numbers.}
    \label{fig:DMD-imaging mapping}
\end{figure}
As a result, we select points at the center of the DMD to ensure that all points will appear in the imaging channel. Keeping this in mind, we display a $4\times4$ array of circles with different radii on to the DMD and then determine their centroids in the imaging channel image. These circles are of different sizes to readily determine the orientation of the imaging channel image of the DMD. 
We rank them by their area as the number labeled beside each circle as shown in Figure~\ref{fig:DMD-imaging mapping}. 
We sort these two arrays of centroids by increasing area of their corresponding circles and then use the ``findHomography" function from the cv2 Python package to calculate the homography. This homography, in the form of a $3\times3$ transformation matrix, relates imaging channel detector coordinates to DMD micro-mirror coordinates as shown in Equation~\ref{eq:matrix}
\begin{equation}\label{eq:matrix}
    \begin{bmatrix}
    x_{DMD} \\
    y_{DMD} \\
    c
    \end{bmatrix}
    = H_{3\times3}
    \begin{bmatrix}
    x_{IMG} \\
    y_{IMG} \\
    1
    \end{bmatrix},
\end{equation}
where $H_{3\times3}$ is the homography, $c$ is a normalization factor, and $x$ and $y$ are the coordinates on each of the two planes. To transform from DMD to Imaging coordinates, we simply multiply $\begin{bmatrix}
    x_{Img} & y_{Img} &1
\end{bmatrix}$ with the inverse of the transformation matrix (i.e. $H_{3\times3}^{-1}$). Using the homography, we are able to actuate the DMD micro-mirrors corresponding to the selected targets as viewed in the imaging channel, directing the light from those targets into the spectral channel.

\subsection{Mapping Result}\label{sec:mapping result}
 
\begin{figure}[htb!]
    \centering
    \includegraphics[width=1\linewidth]{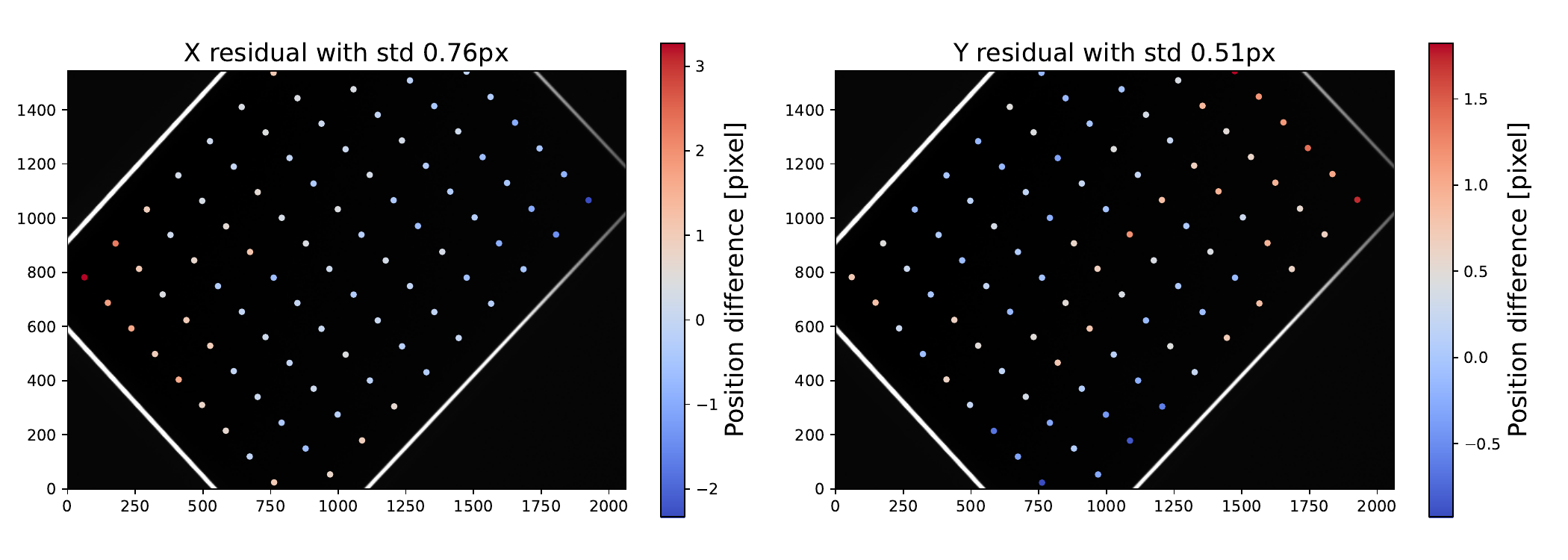}
    \caption{Position different between the centroids from actual image with $10\times10$ pinhole mask and the calculated centroids. Where the different along X-axis is on the left and along Y-axis is on the right. Noticed the white edges are leakage of light from DMD edge. They did not affect the view of the imaging channel, and instead they clearly tell us where are the prohibited region that was not covered by the DMD.}
    \label{fig:DMD-Imag pinhole test}
\end{figure}

We quantify the accuracy of this mapping using a $10\times10$ pinhole grid, with known DMD coordinates for each pinhole. We capture an imaging channel image of this pinhole grid and then calculate the expected positions for each pinhole using the homography. We then determine the difference between the measured and expected positions in both the X and Y directions as we show in Figure~\ref{fig:DMD-Imag pinhole test}.
In the X direction, we observe a range of -2.5 px to 3 px from the expected position with a 0.76 px standard deviation, while in Y direction the range is -1 to 1.7 pixels with a 0.51 px standard deviation. Near the center of the DMD, deviations are minimal, while points at the edges have larger deviations. We attribute this to only using 16 correspondences near the DMD center (see Figure~\ref{fig:DMD-imaging mapping}). Though these results suffice for our purposes, further improvements can be made by generating correspondences further out towards the DMD edges.

\section{Sky test}\label{sec: sky test}

We obtained our first science observations on June 3rd, 2026 at the E.C. Carr Astronomical Observatory (CAO) in Ontario, Canada in 3 arcsecond seeing conditions. The goal of these initial observations was to collect spectra of well studied sources like Vega to check for any systematic issues with DMD-MOS and test our wavelength solution. 

\begin{figure}[htb!]
    \centering
    \includegraphics[width=0.5\linewidth]{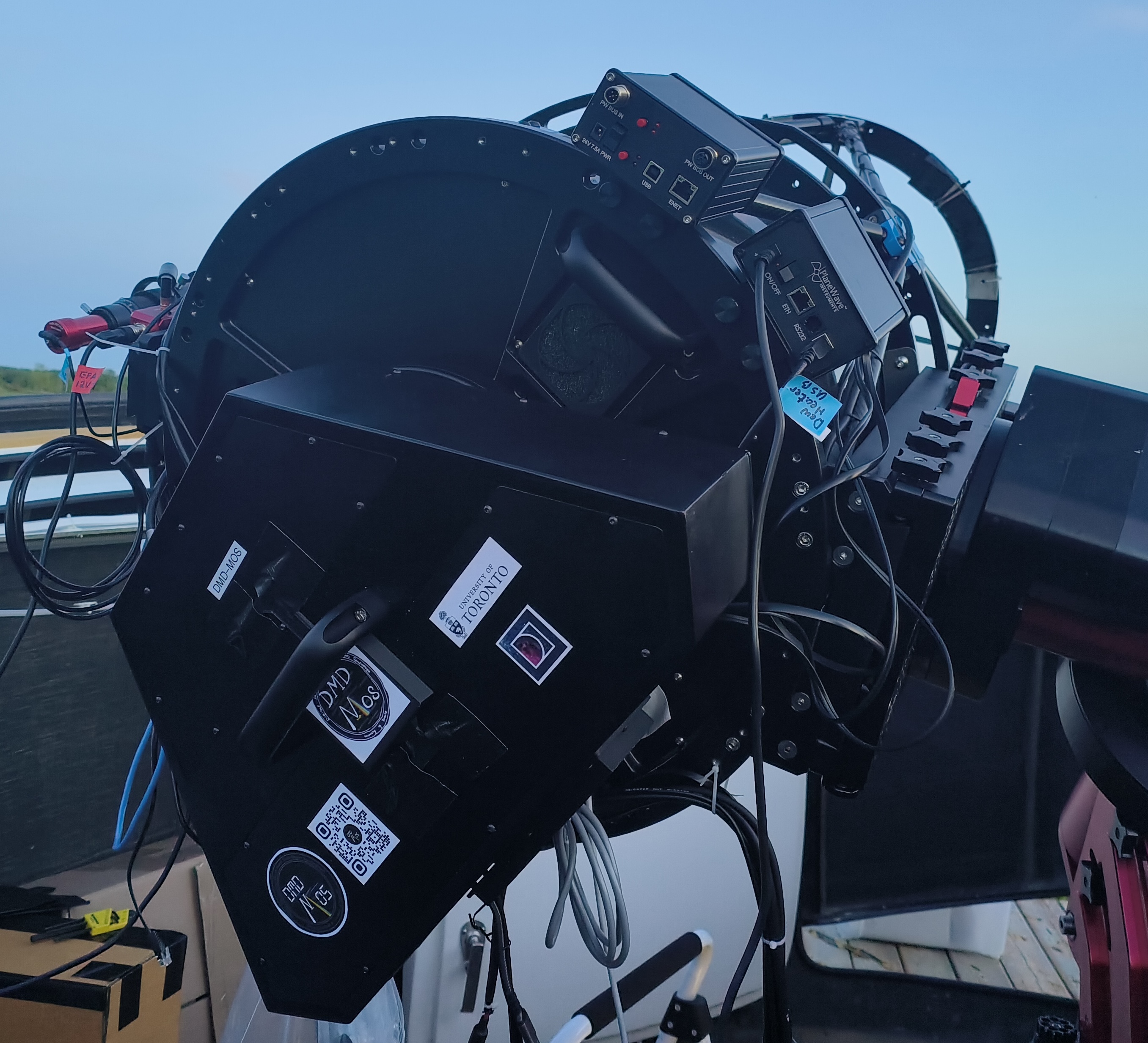}
    \caption{DMD-MOS attached at the back of a CDK20-0.5m telescope at CAO.}
    \label{fig:DMD}
\end{figure}
\begin{figure}[htb!]
    \centering
    \includegraphics[width=0.7\linewidth]{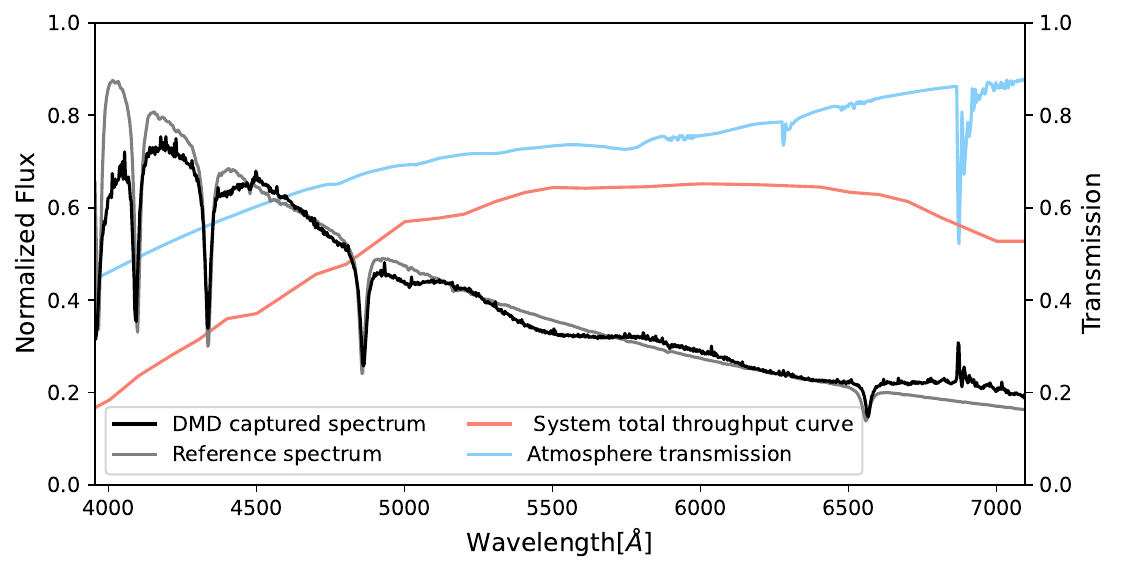}
    \caption{Vega spectrum from DMD-MOS (black curve) compared with a literature Vega spectrum (gray) in the wavelength range of 400-700 nm. The intensities are normalized to approximately the same level for easy comparison. The atmospheric transmission and system throughput are plotted in light blue and red, respectively, with the corresponding Y-axis unit as fractional throughput. The black line is the captured spectrum after reduce atmosphere transmission and total system throughput.}
    \label{fig:Vega spectrum}
\end{figure}

We mounted DMD-MOS onto the CDK20 0.5-meter telescope as shown in Figure~\ref{fig:DMD}. The CDK20 is a f/6.8 Corrected Dall-Kirkham telescope, with a measured FOV of $10.5\times13.4$ arcminutes. For the observations, we used a staircase slit with a $\sqrt{2}$ DMD slit width (similar to the one shown in Figure~\ref{fig:stair case slit}) to collect the spectrum of Vega shown in Figure~\ref{fig:Vega spectrum}.

 We use multiple arc lamps to generate the wavelength solution for our obtained Vega spectrum and compare our data with a reference Vega spectrum from Calspec taken with Hubble Space Telescope data \footnote{\href{https://www.stsci.edu/hst/instrumentation/reference-data-for-calibration-and-tools/astronomical-catalogs/calspec}{\texttt{alpha\_lyr\_stis\_011.fits}} from STScI CALSPEC database.} [\citenum{Bohlin_2020},\citenum{Bohlin2004}]. As the reference spectrum is of a much higher resolution than ours, we convolve it to match the resolution of our system. After scaling both spectra, we see that the absorption lines match fairly well, with less than 2 px deviation; we attribute this deviation to not having more arc lines in the blue end. The overall intensity matches the reference spectrum, but near 400 nm, the DMD-MOS spectrum is significantly lower than the reference curve, which we suspect to be a limitation of the system's throughput and our atmospheric transmission model. For the atmospheric transmission, we adopt the model used in the ESO Sky Calc simulator set to the same resolution as our system with a seeing of 3 degrees for our observation condition. Our total system throughput data includes DMD reflection, transmission in the spectral channel optics, our grating's transmission efficiency, and the quantum efficiency of the detector. DMD-MOS' response peaks at 600 nm with approximately $60\%$ system transmission and drops to less than $20\%$ around 400 nm as the transmission scaling labeled on the right of the figure.

\section{Improvements and Implications}\label{sec: future work}
With the success of our initial on-sky testing, our plan is to complete additional on-sky commissioning tests, including fully testing all functions of the software such as manual target selection and multiple-target acquisition. We plan to collect spectra from different types of stars, stars in clusters, and extended sources (e.g. galaxies) using a DMD long-slit. Beyond upcoming on-sky testing, we look forward to taking advantage of the portable nature of DMD-MOS and testing with larger telescopes, ideally under more favorable observing conditions.

In addition, we will further improve background light levels and ensure that stray light from all sources (e.g. the metal border surrounding the DMD) is thoroughly mitigated. The mapping relation between the DMD and imaging channel detector will also be improved by using additional point correspondences to reduce position differences at the edge of the field. We also aim to improve our wavelength calibration by using arc lamps with more emission lines on the blue end and through better pixel-wavelength matching algorithms in the data reduction pipeline. Future avenues for DMD-MOS include using the imaging channel for photometry, as this is fairly straightforward to include.

Lastly, due to the compact nature of the design, a space-based version of DMD-MOS is an exciting possibility where we can obtain broader wavelength coverage (e.g. near-UV). DMD-based spectrographs could represent a new generation of space-based MOS instruments that can push the boundaries of science.

\section{Conclusion}\label{sec:conclusion}
In conclusion, the Digital Micromirror Device based Multi-Object Spectrograph (DMD-MOS) uses a DMD to instantly create multi-slit masks for the input telescope field, minimizing telescope overhead and facilitating efficient observing for up to hundreds of targets. DMD-MOS has a resolution of 1000 at the center wavelength of 550 nm when using a $2\times2$ micro-mirror slit and wavelength coverage from 400-700nm, with flexible slit controls allow spectrum acquisition for multiple targets under ${10.5}^\prime\times{13.4}^\prime$ FOV. One DMD micro-mirror covers $\sim$0.8 arcsec and is resolved in both imaging and spectral channels with an approximate 2-to-1 ratio between imaging channel detector pixels and DMD micro-mirrors. We use a homography matrix method to relate the imaging channel and DMD planes, which is crucial for accurate slit mask creation for science target observations. During our initial on-sky testing, we collected the spectrum of Vega and can confirm that the obtained DMD-MOS spectrum generally matches literature Vega spectra, providing evidence that DMD-MOS and similar instruments are suitable for ground-based science and beyond.


\appendix    

\acknowledgments 

This work was partially supported by the Dunlap Seed Funding program at the University of Toronto. The Dunlap Institute for Astronomy \& Astrophysics is funded through an endowment established by the David Dunlap family and the University of Toronto. Additionally, this work was also partially supported by the Canada Foundation for Innovation and the Ontario Research Fund.

Jiayi Emma Xu. gratefully acknowledges the SPIE community for the opportunity to present this work, and thanks the full DMD-MOS team, whose collaboration made this instrument possible: PI -- Dr. Shaojie Chen, Dr. Sarik Jeram; Consultants -- Prof. Ting Li, Prof. Suresh Sivanandam; Software -- Zain Azam, Sarah He; System design -- Mark Barnet, Daksh Singh, Coby Silayan; Pipeline testing -- Yixuan (Chris) Cheng; Lab testing -- Ziang (Frank) Chen, Adi Khandelwal, Dhruv Pandya, Laly Boyer; Telescope management -- Prof. Michael Reid. The authors also thank the University of Toronto Department of Astronomy and Astrophysics and the Dunlap Institute for their support.

Emma is grateful to her family members and friends especially to her parents, Xiaoying Chen and Xiang Xu, and to her boyfriend Haosheng Liu, for their support and encouragement throughout this work, as well as her cat Nanxing, for good company along the way.

The cloud will part, the scattering will settle, and the future will resolve into a beautiful spectrum.

\bibliography{report} 
\bibliographystyle{spiebib} 

\end{document}